\documentclass[twocolumn, superscriptaddress,amsmath,amssymb,prb]{revtex4}

\usepackage{amsmath}

\usepackage{graphicx}
\usepackage{subfigure}
\usepackage[english]{babel}

\begin{document}

\title{BN domains included into carbon nanotubes: role of interface}

\author{Viktoria V. Ivanovskaya}
\email{ivanovskaya@lps.u-psud.fr}
\affiliation{Laboratoire de Physique des Solides, Univ. Paris-Sud, CNRS-UMR 8502, 91405, Orsay, France}
\affiliation{Institute of Solid State Chemistry, Ural division of Russian Academy of Science, 620041, Ekaterinburg, Russia}

\author{Alberto Zobelli}
\affiliation{Laboratoire de Physique des Solides, Univ. Paris-Sud, CNRS-UMR 8502, 91405, Orsay, France}

\author{Odile St\'ephan}
\affiliation{Laboratoire de Physique des Solides, Univ. Paris-Sud, CNRS-UMR 8502, 91405, Orsay, France}

\author{Patrick R. Briddon}
\affiliation{School of Electyrical, Electronic and Computer Engineering, University of Newcastle upon Tyne, Newcastle NE1 7RU, United Kingdom}

\author{Christian Colliex}
\affiliation{Laboratoire de Physique des Solides, Univ. Paris-Sud, CNRS-UMR 8502, 91405, Orsay, France}

\begin{abstract}
We present a density functional theory study on the shape and arrangement of small BN domains embedded into single-walled carbon nanotubes. We show a strong tendency for the BN hexagons formation at the simultaneous inclusion of B and N atoms within the walls of carbon nanotubes. The work emphasizes the importance of a correct description of the BN-C frontier. We suggest that BN-C interface will be formed preferentially with the participation of N-C bonds. Thus, we propose a new way of stabilizing the small BN inclusions through the formation of nitrogen terminated borders.
The comparison between the obtained results and the available experimental data on formation of BN plackets within the single walled carbon nanotubes is presented.
The mirror situation of inclusion of carbon plackets within single walled BN nanotubes is considered within the proposed formalism. Finally, we show that the inclusion of small BN plackets inside the CNTs strongly affects the electronic character of the initial systems, opening a band gap. The nitrogen excess in the BN plackets  introduces donor states in the band gap and it might thus result in a promising way for n-doping single walled carbon nanotubes.
\end{abstract}

\maketitle

\section{Introduction}

An effective and efficient way to modify the functional properties of nanomaterials (organic and inorganic nanotubes, fullerenes, nanowires)  is through chemical treatments at- or post- synthesis,   which brings local changes to the original elemental composition.\cite{Func1, Func2} For carbon nanotubes (CNTs) several possible methods have been reported: substitution of carbon atoms by dopant elements, chemi- and physi- sorption of guest atoms (or molecules) outside the tube's walls (side functionalization), and insertion of guest atoms or molecules inside the tubes. \cite{Hirsch, Hirsch-2}
The stability and properties of nanocomposites produced by doping of CNTs are strongly dependent on the nature of dopant atoms.\cite{Papa} In this context, CNTs substitutional doping by nitrogen or boron atoms has been the most studied at the experimental and theoretical levels.\cite{Terrones-Ndoping, Ewels-05, Caroll-98,Blase-99-PRL, Borowiak-03} Spectroscopy studies on nitrogen-doped nanotubes have demonstrated that a  wide range of possible doping concentrations (up to 35 atomic \%) can be obtained by different choices of catalysis and growing conditions.\cite{Trasobares-02} On the contrary, boron doping is found to be rather limited, generally below a few atomic \%.\cite{Gai-04}
The simultaneous inclusion of both boron and nitrogen atoms \cite{BCN-oldest, Terrones-BCN} represents a promising approach for tailoring the electronic properties of nanotubes as a function of their chemical composition. It is well known that pure single walled carbon nanotubes present a metallic, semimetallic or semiconducting behavior depending upon their chiralities. \cite{charlier-07}
In contrast, boron nitride nanotubes are semiconductors with a constant gap of $\sim5.8$ eV independent of their geometries.\cite{Arenal-05,Blase-94,Wirtz-06}
During the past decade, a number of theoretical studies questioning the atomic structure of B-N codoped C tubes have shown that their electronic properties strongly depend on the arrangement and relative concentration of (B,N) over C atoms. Starting from one extreme, several studies have focused on different homogeneous B$_x$C$_y$N$_z$ compounds as BC$_4$N, BC$_2$N or B$_3$C$_2$N$_3$.\cite{Azevedo-06, mazzoni-06,Miyamoto-94, me, Tocite-1, Tocite-2} From the other side a comparative formation energy analysis has clearly demonstrated that the B$_x$C$_y$N$_z$ system tend to demix forming well separated BN and C regions. \cite{blase-another-97,BlaseAlone, blase-99, tomanek-03} The structures proposed correspond to the juxtaposition of BN and C tube sections forming BN-C nanotubes heterojunctions; a large variety of different structural models can be produced by varying the width of the C or BN sections.
Blase et al. proposed that this arrangement gives rise to metal/semiconductor junctions that could be used as Schottky barriers.\cite{blase-another-97} Furthermore, C-BN superlattices can be considered as quantum wells, those electronic properties might be tuned by section widths.
In spite of this high technological potential which has motivated a large number of theoretical works, these BN-C nanotubes heterojunctions have not yet been observed experimentally.
However, the general theoretical result predicting a BN:C demixion  has been confirmed experimentally by various successful attempts to synthesize composite B$_x$C$_y$N$_z$ nanotubes.\cite{Stephan-94,Suenaga-97,Golberg-05,Enouz-08} For instance, it has been shown that a radial segregation can occur in mixed BN-C multi-walled nanotubes leading to independent coaxial BN and C nanotubes.\cite{Suenaga-97} A thorough review on the methods of synthesis, structures and physical properties of these materials have been given by Loiseau\cite{Loiseau-01} or Golberg.\cite{Goldberg-04}

Very recently, Enouz et al.\cite{shaima-07} have reported the synthesis of a new kind of B$_x$C$_y$N$_z$ single-walled nanotubes(SWCNTs) without perfect BN-C demixing. Spatially-resolved EELS (electron energy loss spectroscopy) has suggested the existence of BN plackets with a sub-nanometer diameter (typically made of less than 10 BN hexagons) sequentially distributed along the nanotubes axis. These small plackets are characterized by a high ratio between border and inner atom number. Thus the BN-C frontier might strongly affect the stability and electronic properties of the composite tubes.
However current experimental techniques can not access the exact arrangement and shape of BN inclusions inside single walled CNTs and the structure of the BN-C interfaces.
Complementary structural information can be provided by simulations using \textit{ab initio} techniques.

In this paper we present a density functional theory study of the shape and arrangement of small BN domains embedded into single-walled carbon nanotubes.
In agreement with experiment, we confirm the strong tendency for BN hexagon formation and segregation at the simultaneous inclusion of B and N atoms within the walls of carbon nanotubes, Sec. \ref{sec-hexagons-form}. Previous theoretical works have dealt with extended segregated BN formations of perfect stoichiometry and ``ideal'' interfaces\cite{blase-99}. In Sec. \ref{sec-placket-shape} we focus on the importance of a correct description of the BN-C frontier and we propose a way of stabilizing the small BN inclusions through non stoichiometric borders favoring CN bonds. In Sec. \ref{sec-electronic} we consider the influence of the inclusion of such small BN plackets on the electronic properties of the CNTs. Interestingly, we find that the BN inclusion does not necessarily trigger a metallic to semiconducting transition. Indeed, for the most stable configurations the metallic properties of the initial systems are preserved due to the excess of nitrogen in the structure.
The mirror situation of inclusion of carbon plackets within single walled BN nanotubes is also briefly considered, Sec. \ref{sec-symmetric}.

\section{Computational details}

We performed structural optimizations and electronic structure calculations within the framework of the density functional theory in the local density approximation (DFT-LDA) as implemented in the AIMPRO code.\cite{Aimpro, Aimp-2} All the structures are described within the supercell approach, atomic positions have been optimized using a conjugate gradient scheme. Carbon, nitrogen and boron are described using pseudopotentials which are generated using the Hartwigsen-Goedecker-Hutter scheme.\cite{Hgh-98}
Valence orbitals are represented by a set of Cartesian-Gaussians of s-, p- and d-type basis functions centered at the atomic sites. For carbon, boron and nitrogen we use a large basis set of 22 independent functions. These are based on Gaussians of four different exponents multiplied by the standard Cartesian prefactors. Three of these Gaussians are multiplied by prefactors giving $s-$ and $p-$ type functions (4 functions in total for each exponent); the Gaussian of second smallest exponent is multiplied by Cartesian prefactors to give s-, p- and d-type functions (10 functions in total). The code and these basis sets have been previously successfully applied to study similar BN and graphitic systems. \cite{zobelli, Chris}

A (15,0) semimetallic  zigzag tube has been described in a tetragonal lattice where the cell parameters orthogonal to the tube axis has been chosen in order to have a minimal distance of about 20 \AA \- between adjacent tubes. In order to avoid inter-cell dopant interactions, we chose a supercell (180 atoms) 3 times the elementary nanotube cell along the axis tube direction. A 4 k-point Monkhorst-Pack grid along the tube axis has been used for Brillouin zone integration.
Doping of the graphite layer has been calculated in a $9\times 9\times 1$ supercell with a set of $2\times 2 \times 1$ k-points, the distance between the neighboring layers has been chosen equal to 20 \AA.

In order to compare the relative stability of the considered structures, the formation energies (E$_{form}$) have been estimated as:

\begin{equation}
E_{form} = E(B_xC_yN_z)-(x\mu_B+ y\mu_C+z\mu_N)
\label{formenergy}
\end{equation}

where E$(B_xC_yN_z)$ is the total energy of a mixed system. Dependently upon the studied system (doped graphite or nanotube), the chemical potential for carbon $\mu_C$  has been calculated with respect either to graphite layer or (15,0) carbon nanotube, respectively. The chemical potential $\mu_N$ has been calculated taking the N$_2$ molecule as a reference.
The chemical potential for boron has been calculated as $\mu_B = \mu_{\text{BN}}-\mu_{\text{N}}$, where $\mu_{BN}$ is the chemical potential as derived from BN hexagonal bulk phase.

\section{Results and Discussion}

\subsection{Formation of BN hexagons inside the carbon lattice\label{sec-hexagons-form}}

In order to understand the BN doping of C-nanotubes, we consider firstly the simple case of monoatomic substitution. For simplicity, we start with single boron atom  substitutional doping within a graphene supercell. This substitution is characterized by a formation energy (E$_{\text{form}}$) of  4.26 eV, which lowers up to 3.67 eV within a tube.
The obtained high values of formation energies indicate that a simple substitution of boron within the already formed graphite would be hindered. Indeed, in experimental conditions, the use of a boron-carbon mixed precursor is essential as it helps to the inclusion of boron inside the carbon network.\cite{shaima-07,Gai-04} Nitrogen doping of single walled carbon nanotubes has been discussed in detail in Ref. \cite{Ewels-05}, where it has been proposed that nitrogen substitution  occurs preferentially at vacancy neighboring sites (pyridinic N-vacancy complex). However, even in the simple case of nitrogen replacing a carbon atom,  the formation energies (0.59 eV  for the graphene, 0.62 eV for the tube), are substantially smaller than for B-substitution. In both cases, E$_{\text{form}}$  values indicate an endothermic B or N -substitutional process.

We consider now the simultaneous inclusion of a boron and a nitrogen atom within the graphitic lattice. To estimate the mutual interaction between dopants, we calculate the binding energy as a function of the B-N separation. We discuss the situation when B and N atoms occupy first neighboring sites (i.e. forming a B-N dimer), second neighboring sites and the two inequivalent third neighboring sites.

We define the binding energy E$_{\text{bind}}$ between the B-N dopants as:
\begin{equation}
E_{\text{bind}}=E_{\text{BN}}-(E_{\text{B}}+E_{\text{N}})
\end{equation}
where E$_{\text{BN}}$, E$_{\text{B}}$ and E$_{\text{N}}$ are the formation energies of the systems doped with one BN pair, and single B and N atoms, respectively. Formation and binding energies for different substitutional configurations are presented in Table \ref{tabformenergies}. We obtain the highest binding energy of about 2.64 eV for B and N atoms at first neighboring sites. This indicates that B-N substitution starts preferentially through the formation of BN pairs.

\begin{figure}[tbp]
  \includegraphics[width=\columnwidth]{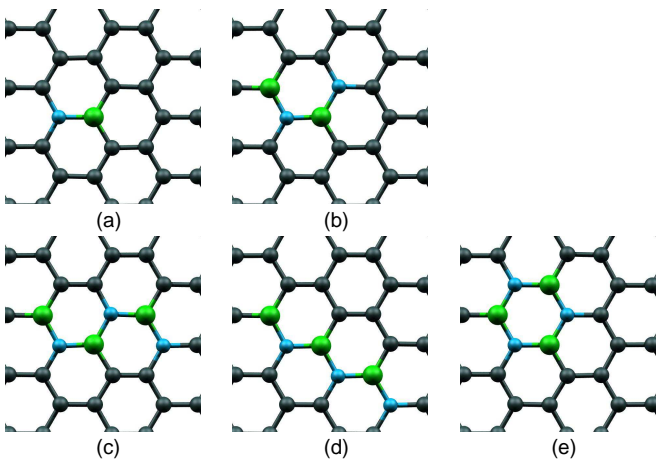}
  \caption{(Color online) Optimized structures for the inclusion of one (a), two (b) and three (c-e) BN pairs within the graphitic layer.}
  \label{plane}
\end{figure}

\begin{table}[b]
\caption{Formation and binding energies for the single atom B and N substitution and for simultaneous inclusion of B and N atoms at the first, second and two inequivalent third neighboring sites within the graphitic plane. }
\label{tabformenergies}
\begin{tabular*}{0.8\linewidth}{@{\extracolsep{\fill}}lcc}
\hline
\hline
& E$_{\text{form}}$ (eV)& E$_{\text{bind}}$ (eV) \\
 \hline
B subs. & 4.26  (3.67)$^1$& \\
N subs. & 0.59  (0.62)$^1$& \\
BN 1$^{\text{st}}(dimer)$ & 2.20  &  2.64 \\
BN 2$^{\text{nd}}$ & 3.47  & 1.38 \\
BN 3$^{\text{rd}}$-1 & 3.72  &  1.13\\
BN 3$^{\text{rd}}$-2 & 3.50
 & 1.35 \\
\hline
\hline

\end{tabular*}

\begin{small}$^1$ in a carbon nanotube\end{small}
\end{table}

The formation energy for the first BN dimer inclusion is as low as 2.20 eV (Fig. \ref{plane}.a). Further addition of a second BN pair, leading to a B-N-B-N line (Fig. \ref{plane}.b), lowers the formation energy per BN unit up to 1.61 eV.
The inclusion of a third BN unit can occur in two different ways: through the BN line or by forming an hexagonal BN-ring. The formation of an hexagon lowers the formation energy per BN unit down to 1.09 eV as compared to the energies of 1.40 eV and 1.53 eV for different linear
configurations (Fig. \ref{plane}, structures c-e).
The obtained results indicate the existence of a strong driving force for the preferential formation of BN hexagonal rings within the graphitic lattice.

\subsection{From single BN hexagons to BN inclusions: ways of ordering, plackets shape and role of the interface}\label{sec-placket-shape}

In the previous section we have seen how the successive inclusion of BN units in the carbon lattice leads preferentially to the formation of hexagons instead of open BN chains.
At higher doping levels, we should take into account the ordering of the BN hexagons, which might occur in two different ways: either a random distribution or clustering into BN patches within the walls of the original tube.

We begin by considering the case of random ordering of BN hexagons, and we analyze the effect of inter-hexagon interactions on the stability of the system as a function of the hexagon separation.
We choose two orientations for the BN hexagon pairs: translated (T), where the two hexagons can be superposed by a simple translation, and rototranslated (RT), when the two hexagons can be superposed by a translation followed by a 60 degree rotation of one of the rings.
When two hexagons occupy second neighboring hexagonal sites, the formation energies for the T and RT configurations are equal to 6.89 eV  and 6.92 eV, respectively. This small energy difference indicates a lack of selectivity for the relative orientation of the hexagons.
Successively, we increase the BN hexagon's separation by including one or more full carbon hexagonal rings in between. This varies the formation energies of both T and RT systems by a few hundredths of an eV. Thus, the mutual interaction between BN hexagons might be considered as negligible already at second neighboring hexagonal sites.

\begin{figure}[tbp]
\subfigure[E$_f$=9.08 eV]{\includegraphics[width=0.3\columnwidth]{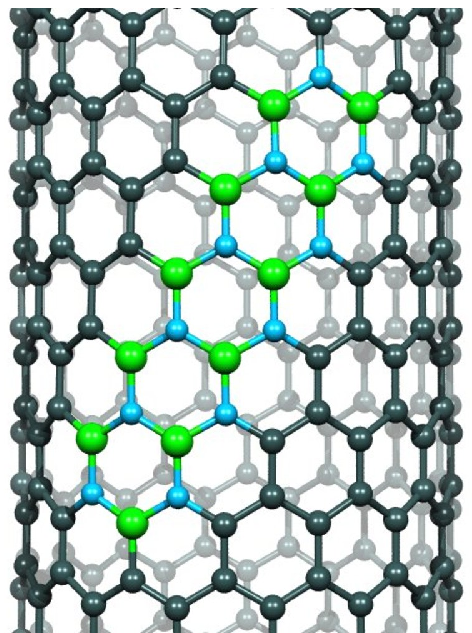}}
\subfigure[E$_f$=8.38 eV]{\includegraphics[width=0.3\columnwidth]{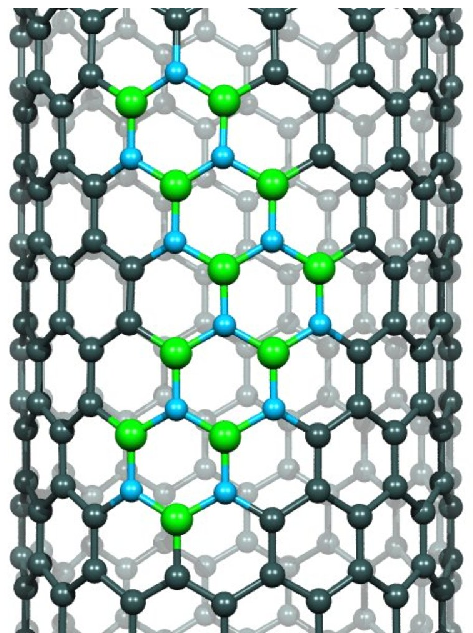}}
\subfigure[E$_f$=8.04 eV]{\includegraphics[width=0.3\columnwidth]{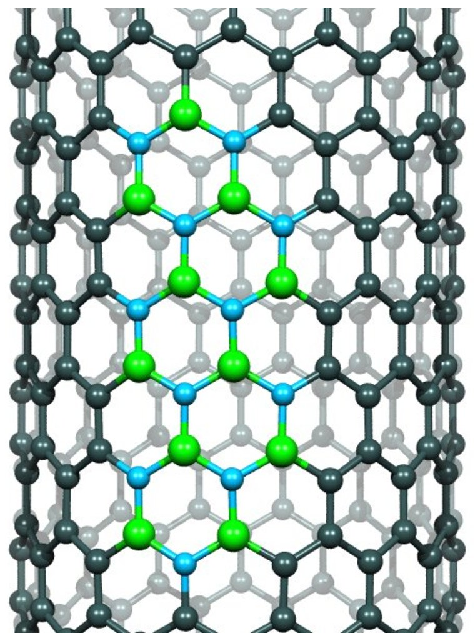}}

\subfigure[E$_f$=7.29 eV]{\includegraphics[width=0.3\columnwidth]{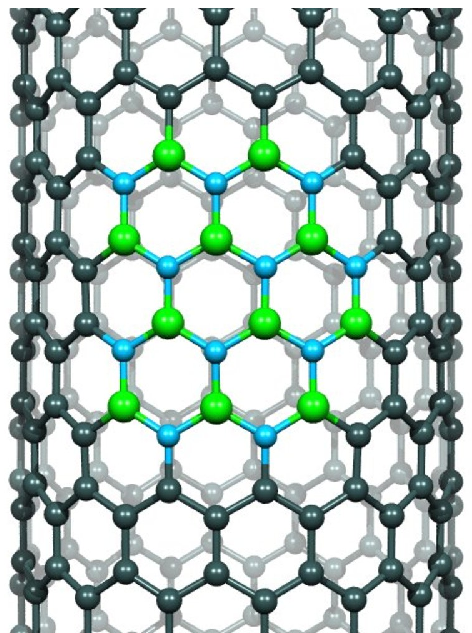}}
\subfigure[E$_f$=7.64 eV]{\includegraphics[width=0.3\columnwidth]{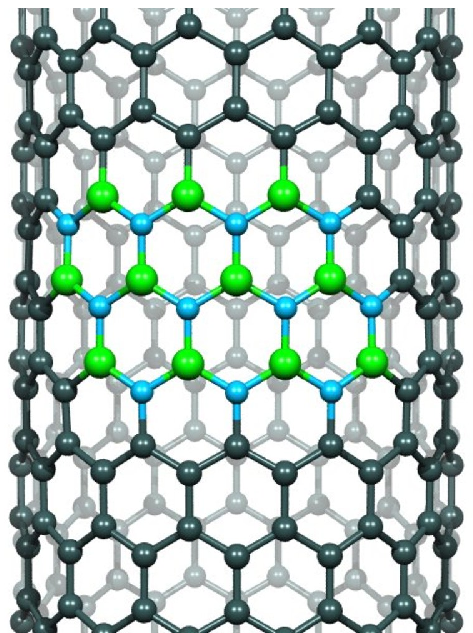}}
  \caption{(Color online) Optimized structures and formation energies for BN inclusions, forming linear hexagonal motifs (a-c) or compact plackets(d-e) within the (15,0) carbon nanotube. The atomic composition is kept identical.}
  \label{linepath}
\end{figure}

We discuss now the clustering of BN hexagons, i.e. the BN segregation within the tube walls in the form of domains. We chose several possible configurations, such as linear hexagonal motifs (Fig. \ref{linepath}, a-c) or more compact BN plackets (Fig. \ref{linepath}, d-e). The overall atomic composition and  patch's stoichiometry (B/N$=$1) were kept identical. The E$_{form}$ values depicted at Fig. \ref{linepath} indicate that the compact hexagon arrangements are energetically the most accessible.

\begin{figure}[tbp]
  \includegraphics[angle=-90,width=\columnwidth]{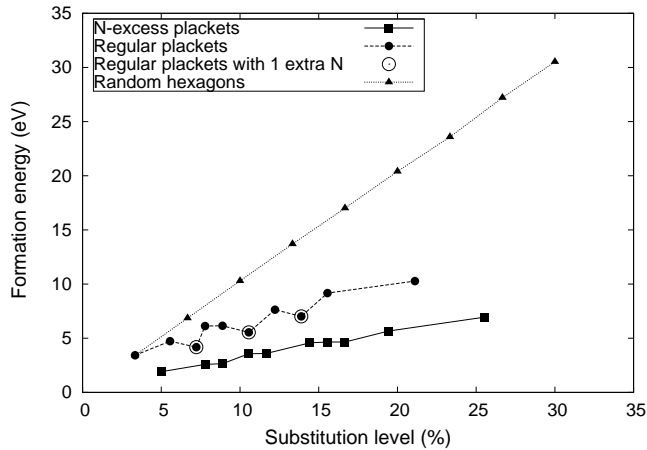}
  \caption{Formation energies for the BN inclusions as a function of their concentration within the tube. Triangles, circles and squares denote respectively a random BN hexagons distribution, regular BN plackets and N excess BN plackets.}
  \label{graphFormen}
\end{figure}

Finally, we compare the stabilites of the stable configurations of random and compact hexagons orderings. Fig. \ref{graphFormen} presents the formation energies of the mixed systems as a function of the BN substitution.
The absence of inter-BN hexagons interactions in the case of separated hexagonal rings leads to a monotonic linear increase of the formation energies with increasing of the dopant content within the CNT.
However, these formation energies are systematically higher than those for the BN plackets over the whole range of the concentrations considered. In particular, this implies a preference for the formation of compact extended BN patches over a random distribution of single BN hexagons within the walls of the carbon tubes.
This trend can be explained by simple considerations based on the lower number of C-B and C-N bonds within the compact structures compared to the separated BN hexagons or to the linear BN chains of the equal compositions, see Fig. \ref{linepath}.
Indeed, the well known hierarchy of the energetic stability of the bonds [B-N $>$ C-C $>$ C-N $>$ C-B]\cite{nozaki-96} suggests that the most stable system would 1. maximize the number of B-N and C-C bonds\cite{nozakiDR-96,Azevedo-06,azevedoLayer-06b, Tocite-1, Tocite-2}and 2. form preferentially C-N bonds over C-B ones. Thus, the condition 1 is satisfied by the formation of compact configurations.

\begin{figure}[bt]
\subfigure[B$_3$N$_3$]{\includegraphics[width=0.45\columnwidth]{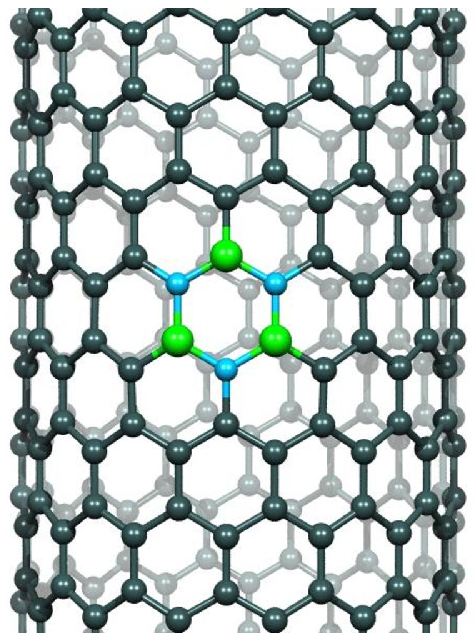}\label{B3N3-struct}}
\subfigure[N terminated B$_3$N$_6$]{\includegraphics[width=0.45\columnwidth]{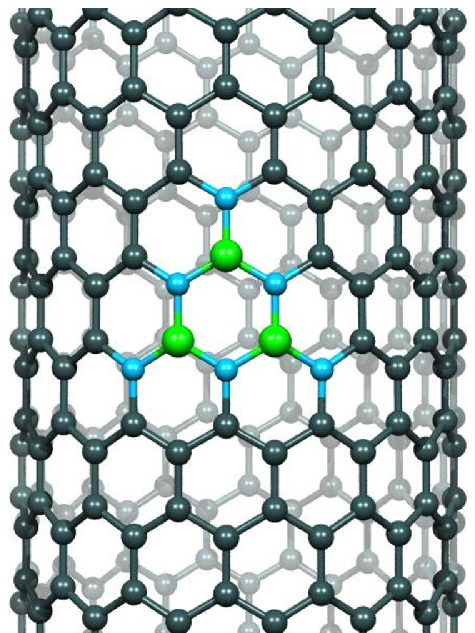}\label{B3N6-struct}}

\subfigure[N atom excess B$_9$N$_{10}$]{\includegraphics[width=0.3\columnwidth]{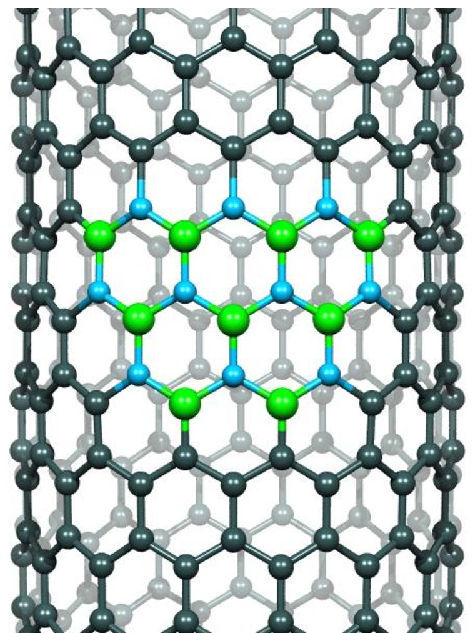}\label{B9N10-struct}}
\subfigure[B atom excess B$_{10}$N$_9$]{\includegraphics[width=0.3\columnwidth]{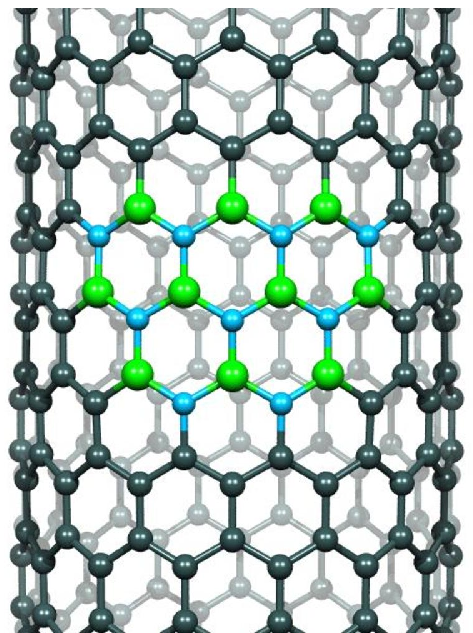}\label{B10N9-struct}}
\subfigure[N terminated B$_{10}$N$_{16}$]{\includegraphics[width=0.3\columnwidth]{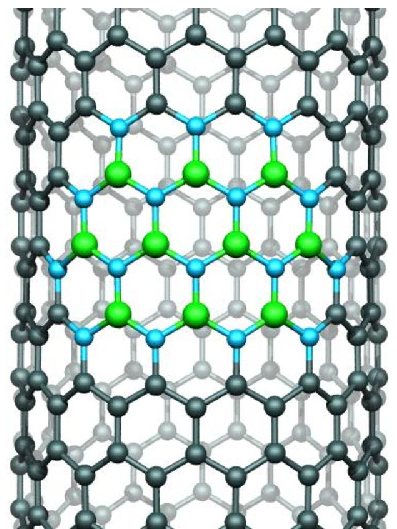}\label{B10N16-struct}}
\caption{(Color online) Optimized structures for complete (a), one-atom-excess (c,d) and nitrogen terminated (b, e) BN plackets within CNT.}
\label{graphFormen2}
\end{figure}

Hitherto, all the structures that have been considered for the composite tubes have had an equal number of boron and nitrogen atoms. However, certain BN domain morphologies, i.e. odd BN hexagons arranged in non-linear patches, may answer to both conditions (1) and (2) through an excess of one nitrogen atom, see Fig.\ref{graphFormen2}(c). Formation energies for the domains with one extra nitrogen are presented in Fig.\ref{graphFormen} and are about 1.5 eV lower than for systems with matching numbers of B and N atoms. The formation energy for a B$_9$N$_{10}$ domain presented in Fig.\ref{graphFormen2} (c) is about 3.4 eV lower than for an analogous structure with one odd boron atom (d).

Further increases of the nitrogen excess in the system  may lead to full exclusion of the C-B bonds at the placket/tube interface. The idea of BN/C heterostructures has been earlier proposed by Blase et al. \cite{blase-99, blase-another-97, BlaseAlone}  Simulating the process of interdiffusion at the C/BN interface by exchanging C with N or B atoms, the authors found the formation of abrupt C/BN interfaces to be energetically favorable.
However, the interfaces proposed did not fully satisfy the ``stability criteria'' given above and contained all possible types of bonds, including C-B bonds. A further step is to consider the C/BN interfaces decorated solely by C-N bonds.

As an example, in Fig. \ref{graphFormen2} (b and e) we present new models for the B$_x$N$_y$ (x $<$ y) inclusions, which do not preserve the conventional (B/N$=$1) placket's stoichiometry. In the composite tube's structure all energetically non favorable interfacial C-N bonds have been eliminated, while the frontier is formed just by C-N bonds. We thereby obtain the result that, over the whole range of concentrations considered, the formation of such nitrogen excess B$_x$N$_y$ domains leads to the lowest formation energies. The energy gain compared to the regular plackets measures up to several eV, see Fig. \ref{graphFormen}. The B/N ratio increases gradually with the domain's size: at the lower limit of concentrations studied (5 \%) the ratio is equal to 0.5, while at the upper one ($\sim 26$ \%) it is increased to 0.7.

The proposed model agrees well with recent experimental findings on BN nanodomains inside single-walled CNTs.\cite{shaima-07} Enouz et al. have shown that fine structures for B-K edges at the BN-C heterojunctions  do not show any significant presence of C-B bonds, in contrast to those for N-K edge that suggest a non negligible concentration of C-N bonds.
This indicates that the boundaries between the BN domains and the C network are dominated by C-N bonds\cite{shaima-07}. This preferential tendency is consistent with the greater ease with which nitrogen atoms can be incorporated inside carbon nanotubes as compared to boron atoms. The phenomenological model based on energetic considerations and on the phase separation between graphite and h-BN is described in detail elsewhere. \cite{shaima-07}

\subsection{Symmetric situation: C domains embedded in BN nanotubes}\label{sec-symmetric}

As an extension of the scenario illustrated above, here we briefly discuss the mirror situation of carbon doping  BN nanotubes.
Several theoretical studies on doping BN NTs have considered carbon substitution of single boron or nitrogen\cite{blase-99,wu-05,wu-Another5, Li-CdopedBN} atoms or BN atom pairs\cite{Guo}. It has been shown that doping carbon into the BN structure introduces additional electronic levels within the electronic band gap localized at the impurity site.\cite{wu-05}
However there is no experimental evidence for C-doped BN nanotubes, even if a substitutional reaction has been proposed as a possible method for their synthesis.\cite{terrones-07}

In analogy with the structural models presented before, we have considered a random distribution of C hexagons and compact carbon domains embedded in BN nanotubes. Both regular plackets  and plackets with exclusively C-N bonds at the frontiers have been taken into account. In the case of regular plackets the same number of B and N atoms have been substituted, in the other case the system presents a N excess due to the removal an higher number of B atoms.

We obtain the same trends as discussed above for BN domains in C nanotubes with the lowest formation energies correspond to carbon domains solely framed by C-N bonds. Comparing to the case of BN doping within CNTs,  we get  a remarkable but expected difference in the values of  E$_{\text{form}}$ (about 7 eV more, independently of the substitution level) indicating that BN substitution inside the carbon NTs would be a more energetically feasible process.
Indeed, it is not energetically favorable to substitute the more stable B-N bonds by C-C bonds. These results may explain why C-doped BN nanotubes have not yet been experimentally observed and may indicate that very demanding conditions are required for their synthesis.

\subsection{Electronic properties of BN/C composites}\label{sec-electronic}

In this section we present the modifications induced on the electronic structure of SWCNTs due to the introduction of BN plackets.

\begin{figure*}[tbp]
\subfigure[(15,0) pure carbon tube]{\includegraphics[width=0.3\textwidth]{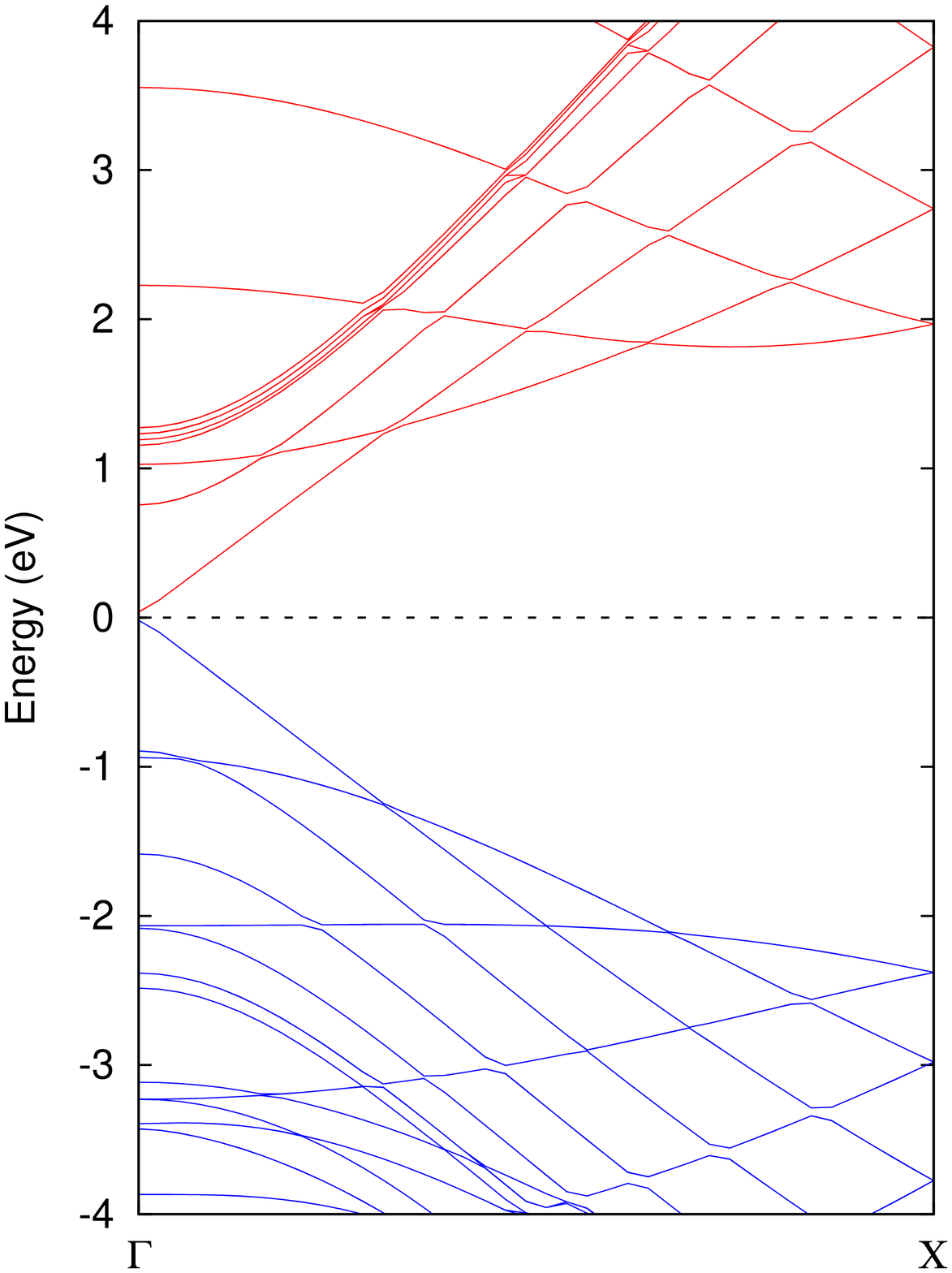}\label{band-pure}}
\subfigure[B$_3$N$_3$ placket within CNT, doping level $\sim$ 3 \%]{\includegraphics[width=0.3\textwidth]{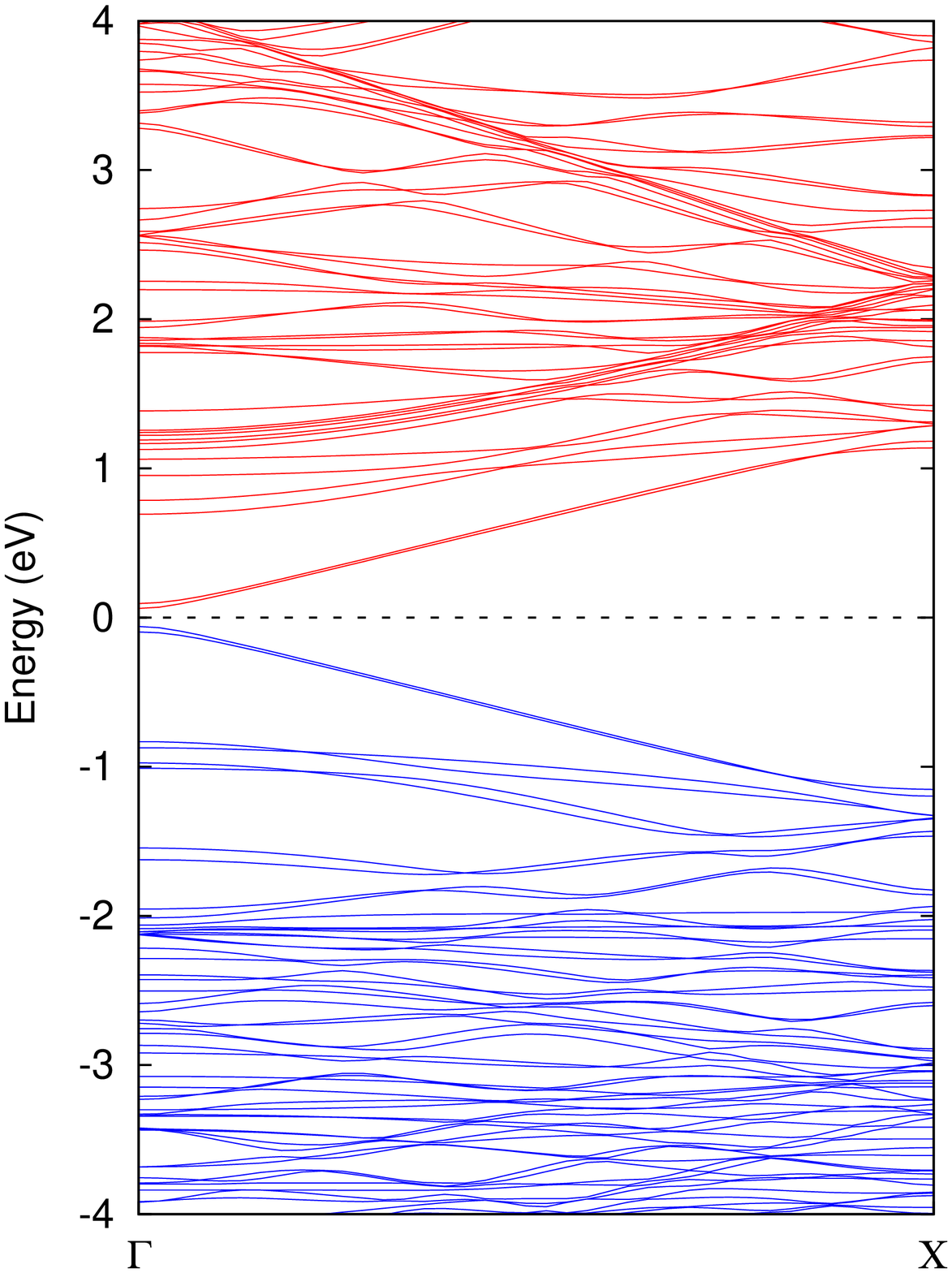}\label{band-1ringstoich}}
\subfigure[B$_{14}$N$_{14}$ placket within CNT, doping level $\sim$ 16 \%]{\includegraphics[width=0.3\textwidth]{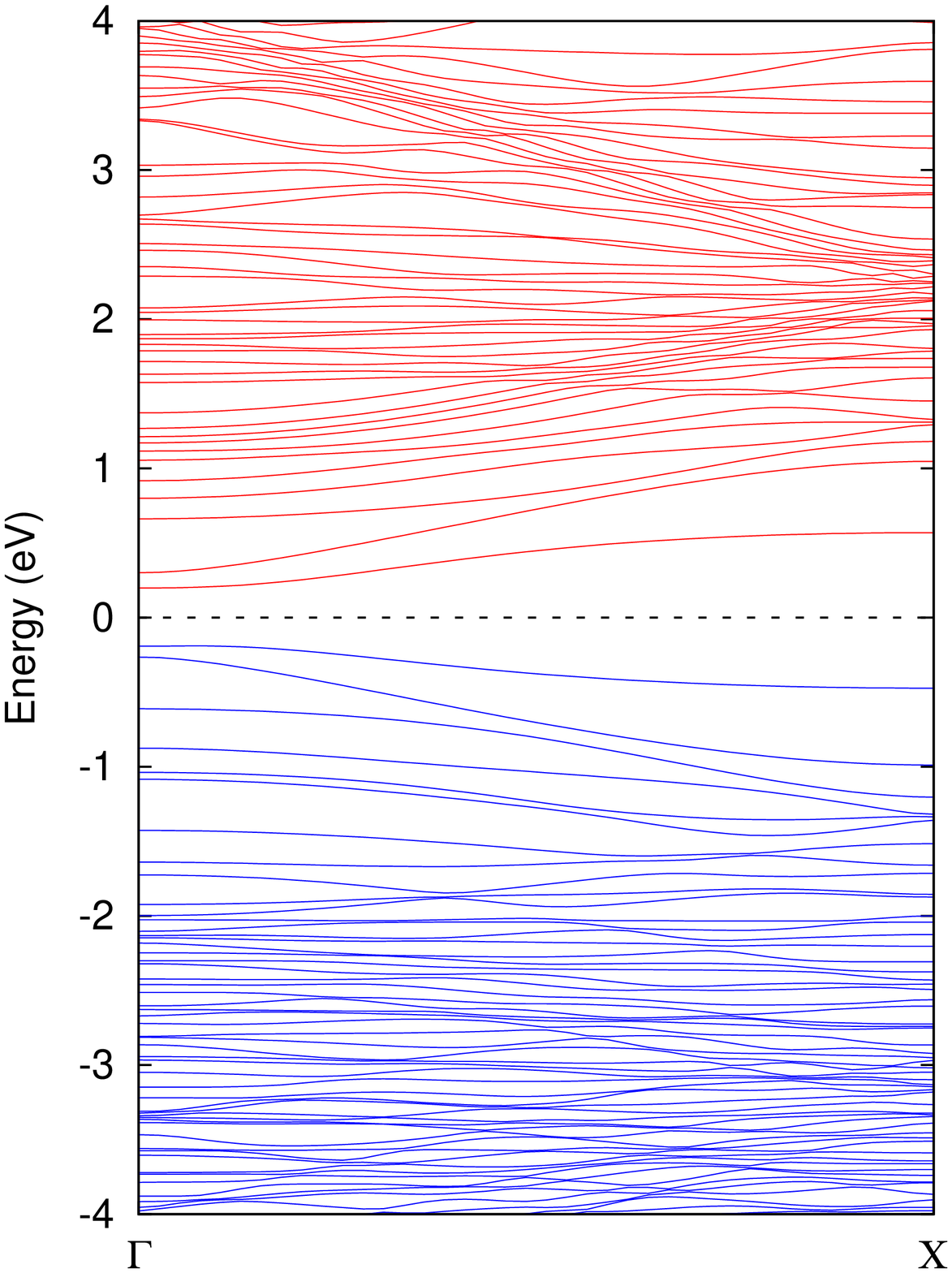}\label{band-1largestoich}}

\subfigure[N-excess B$_3$N$_6$ placket within CNT, doping level $\sim$ 5 \%]{\includegraphics[width=0.3\textwidth]{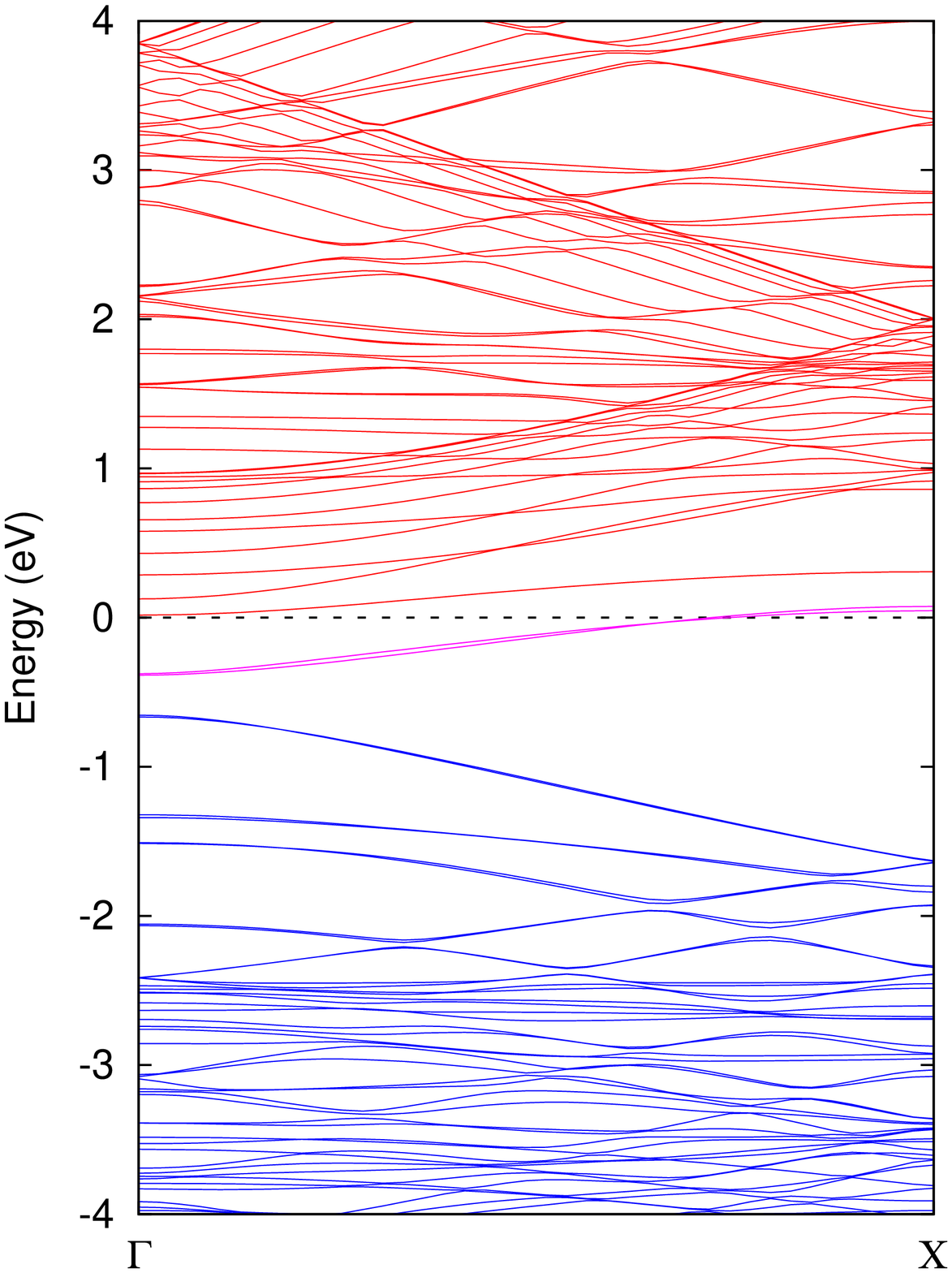}\label{band-1ringNstoich}}
\subfigure[N-excess B$_{11}$N$_{17}$ placket  within CNT, doping level $\sim$ 16 \%]{\includegraphics[width=0.3\textwidth]{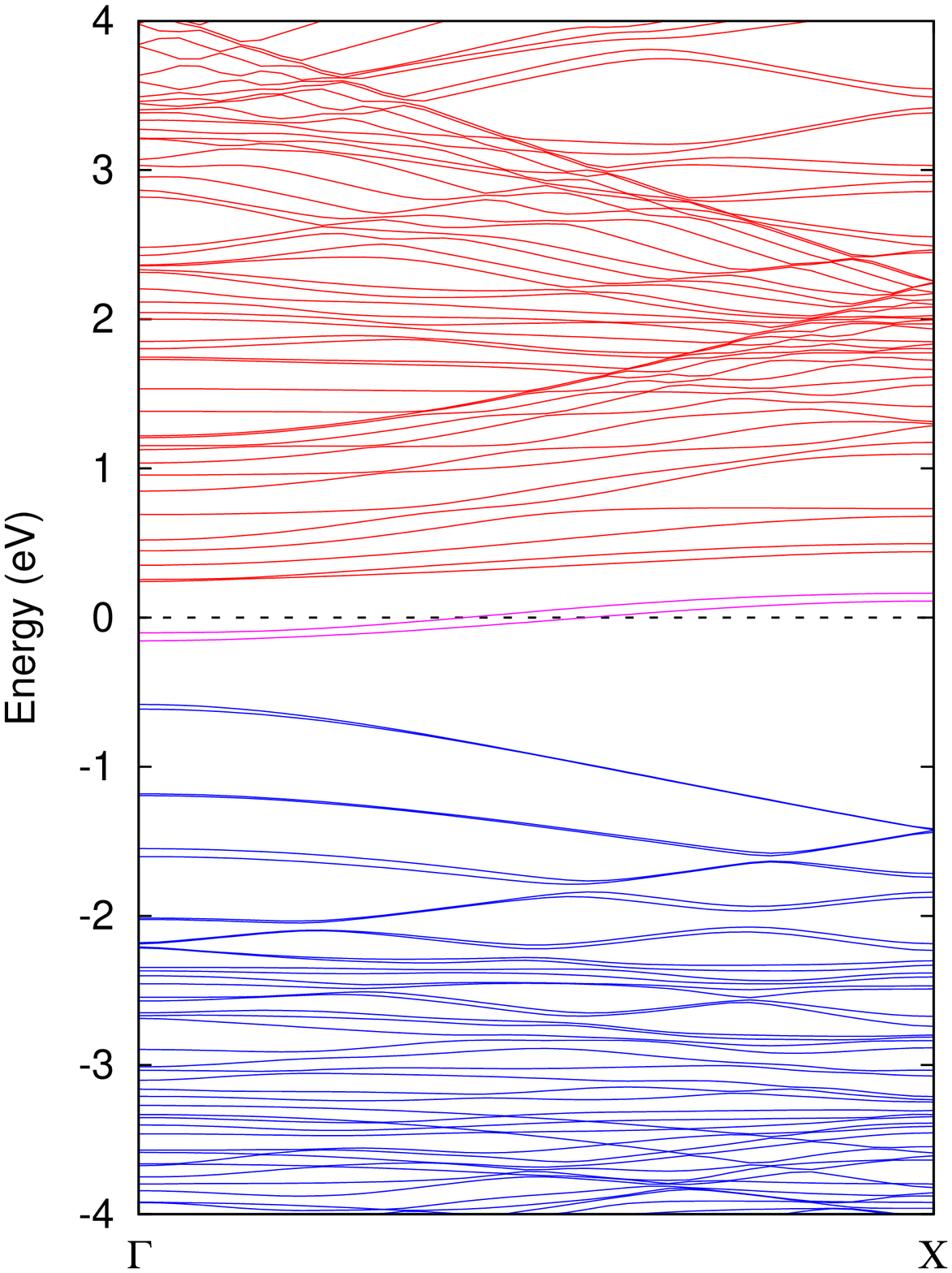}\label{band-1largeNstoich}}
\caption{(Color online) Near Fermi band structures for the pure (a) and doped by BN patches carbon nanotubes (b-e). Occupied bands are plotted in blue, unoccupied in red and bands cut by the Fermi level in magenta (dotted line).}
\label{FigBands}
\end{figure*}

The electronic band structure of a (15,0) semimetallic carbon nanotube is presented in Fig. \ref{band-pure}, where occupied bands are represented in blue and unoccupied in red.
First, we consider the domains where the B over N ratio is kept equal to one (B$_x$N$_x$); in this case the composite doped tube is isoelectronic with the initial carbon nanotube. The inclusion of one B$_3$N$_3$ ring, which corresponds to about 3 \% of substitution within the chosen supercell, leads to the opening of a small band gap of 0.12 eV, see Fig. \ref{band-1ringstoich}. When the BN content is increased up to about 16\%, the electronic gap rises up to 0.39 eV together with a reduction of the band's dispersion due to the higher dopant concentration, see Fig. \ref{band-1largestoich}.
Thus, the inclusion of B$_x$N$_x$ domains leads to a semimetal-semiconductor transition where the electronic band gap can be tuned by the BN placket's size.
This behavior is analogous with that obtained for BN-C nanotube heterojunctions where the band gap depends on the BN section width.\cite{blase-another-97}

However, in Sec. \ref{sec-placket-shape} we saw that composites with nitrogen saturated interfaces are the most energetically favorable systems. Compared with the case of B$_x$N$_x$ domains, doping by nitrogen saturated plackets (B$_x$N$_y$, $x<y$) will not be isoelectronic with the perfect carbon nanotube. We consider now a nitrogen excess B$_3$N$_3$ placket, i.e. B$_3$N$_6$, that is about 5\% of substitution, see Fig.\ref{graphFormen2}. 
For it, as in the case of B$_x$N$_x$ plackets, a small band gap is opened. However, three additional electrons which are introduced into the system by the nitrogen atoms excess lead to the occurring of additional electronic levels within the band gap about 0.3 eV below the bottom of the conduction band (represented in magenta in Fig. \ref{band-1ringNstoich}). A similar behavior is observed as well for a bigger domain B$_{11}$N$_{17}$ (16\% of doping), see Fig.\ref{band-1largeNstoich}; as for the case of stoechimetric domains, the induced band gap increases with the BN content.

Thus, for non stoechiometric domains the mixed nanotubes are $n$-doped. A similar behaviour has been shown in single nitrogen atom doping of zig-zag carbon nanotubes where additional electronic states are localised in a fairly large region (about 30 \AA) around the impurity site.\cite{Nevidomskyy-03}\footnote{In our case, we obtain a similar large wavefunction extension and the overlap of the impurity orbitals in the periodic structure generates an impurity band rather then a set of discreat levels corresponding to the weak doping limit. Tests conducted doubling the supercell size show no influence on the domain formation energies whereas the dispersion of the additional bands in the electronic band gap is significantly reduced.}

\section{Conclusions}

In the present paper we have investigated, using the density functional theory formalism, the structure and electronic properties of the recently synthesized hybrid systems with small BN domains included into single-walled carbon nanotubes.
At the simultaneous inclusion of B and N atoms within the nanotubes wall, we have shown a strong tendency for the formation  of BN pairs and subsequently BN hexagons within the carbon lattice. Particular attention has been paid to the shape and arrangement of BN inclusions.  At higher doping concentrations we found that formation of compact BN domains would be the most energetically favorable.

Furthermore, we have emphasized the importance of a correct description of the BN/C frontier, due to the large border over surface ratio of small dopant domains. Compared to previous studies, we have considered that BN plackets of non ideal stoichiometry might be generated during the synthesis. In particular, we suggest that a nitrogen excess and the formation of nitrogen terminated borders can stabilize these small BN inclusions. This effect can be explained by the sole presence of N-C bonds at the BN/C interface and the absence of the least energetically stable C-B bonds. This result is in agreement with first available experimental data and might be general to any extended BN-C layered structures.

Moreover, we briefly described within the proposed formalism the mirror situation of the inclusion of carbon plackets within single walled BN nanotubes. On the basis of energetic considerations, we suggest that C-domain formation in BN nanotubes can hardly be obtained and might demand very particular synthesis conditions.

In accordance to previous works, we have shown that the inclusion of small BN plackets inside the CNTs strongly affects the electronic character of the initial semimetallic system with the opening of an electronic band gap. However we demonstrate that the inclusion of N-excess BN structures, most energetically stable, introduce additional donor levels within the band gap.

This behaviour is analogous to $n$-doping of single walled carbon nanotubes by individual nitrogen atoms. Heteroatom doping of carbon nanotubes has received considerable attention in the field of nanoelectronic and the generation of localised electronic states is also a promising way for controlled chemical functionalisation of the tubes.\cite{Nevidomskyy-03} However it is worth to notice that actual synthesis 
methods limit the doping concentration to around 1\%\cite{Glerup-04,Lin-09} and that nitrogen dopants are mostly associated with topological defects as vacancies.\cite{Ewels-05}
The formation of BN domains in carbon nanotubes can thus be seen as an alternative and more effective way of doping which might introduce higher donor concentration as well as minor perturbations of the nanotube structure.

The new generation of aberation corrected transmission electron microscopes is now able to provide spectroscopic information at the single atom level.\cite{Muller-08} Atomically spatially resolved EELS experiments are currently in progress on the systems discussed here in order to provide the stoichiometry and local chemical environment. Furthermore, the electronic structure modifications induced by the BN doping might be investigated by means of scanning transmission spectroscopy. 

\section{Acknowledgement}

V.V.I. thanks the Foundation of the President of Russian Federation (Grant -502.2008.3) for financial support. The authors would like to acknowledge C.P. Ewels for inspiring discussions.

\end{document}